\documentstyle[12pt]{article}
\begin{document}
%preprint numbers to be removed for non-preprint version
%from here ...
\begin{flushright}
KUL-TF-95/34 \\
NIKHEF 95-060 \\
hep-th/9510176 \\
\end{flushright}
\vspace{-1.2cm}
%         ... to here

%\renewcommand{\theequation}{\thesection.\arabic{equation}}
\newcommand{\eqn}[1]{eq.(\ref{#1})}

\renewcommand{\section}[1]{\addtocounter{section}{1}
\vspace{5mm} \par \noindent
  {\bf \thesection . #1}\setcounter{subsection}{0}
  \par
   \vspace{2mm} } %was 5mm
\newcommand{\sectionsub}[1]{\addtocounter{section}{1}
\vspace{5mm} \par \noindent
  {\bf \thesection . #1}\setcounter{subsection}{0}\par}
\renewcommand{\subsection}[1]{\addtocounter{subsection}{1}
\vspace{2.5mm}\par\noindent {\em \thesubsection . #1}\par
 \vspace{0.5mm} }
\renewcommand{\thebibliography}[1]{ {\vspace{5mm}\par \noindent{\bf
References}\par \vspace{2mm}}
\list
 {\arabic{enumi}.}{\settowidth\labelwidth{[#1]}\leftmargin\labelwidth
 \advance\leftmargin\labelsep\addtolength{\topsep}{-4em}
 \usecounter{enumi}}
 \def\newblock{\hskip .11em plus .33em minus .07em}
 \sloppy\clubpenalty4000\widowpenalty4000
 \sfcode`\.=1000\relax \setlength{\itemsep}{-0.4em} }

\vspace{4mm}
\begin{center}
{\bf Redefining B-twisted topological sigma models}\footnote{Talk
given by S. Vandoren} \vspace{1.4cm}

F. DE JONGHE$^{a}$, P. TERMONIA$^{b}$, W. TROOST$^{b}$
 and S. VANDOREN$^{b}$ \\[.7cm]
{\em $^{a}$ NIKHEF-H },
{\em Postbus 41882},\\{\em 1009 DB Amsterdam, The Netherlands}\\
{\em $^{b}$Instituut voor Theoretische Fysica, K.U.Leuven} \\
{\em Celestijnenlaan 200D, B-3001 Leuven, Belgium} \\
\end{center}

\centerline{ABSTRACT}
%\vspace{- 4 mm}  %\end{center}
\begin{quote}\small
The recently proposed procedure \cite{tlgqaq} to perform the topological
B-twist in rigid $N=2$ models is applied to the case of the $\sigma $
model on a K\"ahler manifold. This leads to an alternative description of
Witten's topological $\sigma $ model, which allows for a proper BRST
interpretation and ghost number assignement. We also show that the
auxiliary fields, which are responsible for the off shell closure of the
$N=2$ algebra, play an important role in our construction.
\end{quote}
\addtocounter{section}{1}
\par \noindent
  {\bf \thesection . Introduction}
  \par
   \vspace{2mm} %was 5 mm

\noindent

{}From the physical point of view, topological field theories (TFT)
\cite{tqft,tsm} are interesting because they describe certain apects of
$N=2$ or $N=4$ models. They can be solved exactly since the semi-classical
approximation for these theories is exact. For more on motivation and
introduction, see the contribution of R. Dijkgraaf to this volume.

Topological field theories are
field theories % with a BRST symmetry, and
whose energy-momentum tensor is
BRST exact. Formally this implies, via the Ward identity, that the
partition function of the theory is independent of the metric on the
manifold on which the theory is defined. A large class of TFT's can be
constructed by gauge fixing a topological invariant
\cite{gftft} or by the so-called {\it twisting} $N=2$ theories with
\cite{tqft,tsm,Twist}. This
twisting, in turn, can be done in two
different ways,  the so-called A- and B-twist \cite{ABTwist,BF}.

In this paper we consider the twisting of two dimensional $N=2$ $\sigma $
models. These twists both involve changes in the  spins of the fermionic
fields, and the choice of a BRST operator, with the help of
the susy charges of one of the two the $N=2$ algebras \cite{ABTwist,tsm}.
The relevant physical operators (observables) are representatives of the
BRST cohomology classes at some definite ghost number.

The assignment of these ghost numbers and the BRST interpretation for the
A-twist is
straightforward, but for the B-twist it is unclear.
If we do not introduce auxiliary fields in
the $N=2$ algebra, the action after the B-twist
does not have the structure of a gauge fixed action
of an underlying gauge theory. If we do introduce
them, it is the interpretation of the BRST charge itself which is doubtful.

In this contribution we intend to show that, by rewriting the customary
BRST charge for the B-twisted model as the sum of a new BRST charge
and an anti-BRST
charge, the ghost number assignments and the BRST interpretation fall into
place. This procedure can be applied to any (rigid) $N=2$ theory.
For topological Landau-Ginzburg models, see \cite{tlgqaq}.

We propose to take
for the BRST operator $one$ of the N=2 supersymmetry charges used by
Witten, and the other as the anti-BRST operator. The corresponding
ghost number assignments
make a conventional separation in classical fields, ghosts
and antighosts straightforward, but the usual symmetry between BRST and
anti-BRST transformations is not present yet. The interpretation of the
anti-BRST transformation takes an entirely standard form, if one changes
to a different basis of fields, which is related in a (mildly) nonlocal
way with the customary basis.

As a
consequence of our procedure, the (++) component of the energy momentum
tensor is anti-BRST exact while the $(--)$ component is BRST exact. This
implies that
we also need the Ward identity for the anti-BRST operator in order to
prove that the theory is metric independent. Moreover, it also implies that
observables are subjected to two conditions, namely they should be BRST
invariant and their anti-BRST transformation should be BRST exact. This
leads us to define the physical spectrum as being the elements of the
anti-BRST cohomology defined in the BRST cohomology. We argue that this
cohomology problem leads to the same observables as in the old approach.

\section{N=2 $\sigma $-models}
\noindent
One can formulate $N=2$ models with or without auxiliary
fields. Including these fields, one realises the algebra off shell.
We will treat here both cases and comment on the difference between
the two B-twists.
\subsection{On shell formulation}
\noindent
The N=2 $\sigma $-model action on the target K\"ahler manifold ${\cal
M}_K$ is \begin{eqnarray}
S&=&-\frac{1}{2}g_{ij^*}(\partial _+X^i\partial _-X^{j^*}+\partial
_-X^i\partial _+X^{j^*})\nonumber\\
&&+ig_{ij^*}(\psi ^i\nabla _-\psi ^{j^*}+\psi ^{j^*}\nabla _-\psi
^i)\nonumber\\
&&+ig_{ij^*}(\xi ^i\nabla _+\xi ^{j^*}+\xi ^{j^*}\nabla _+\xi
^i)\nonumber\\
&&+4R_{ij^*kl^*}\psi ^i\psi ^{j^*}\xi ^k\xi ^{l^*}\ ,
\end{eqnarray}
where e.g. $\nabla _-\psi ^i=\partial _-\psi ^i-\Gamma ^i_{jk}\partial
_-X^j\psi ^k$, and an integral over a Riemann surface $\Sigma $ is also
understood. The supersymmetry rules are
\begin{equation}
\begin{array}{cc}
\delta X^i  =  \psi ^i \epsilon ^- + \xi ^i {\tilde \epsilon}^- &
\delta X^{i^*} = -\psi ^{i^*} \epsilon ^+ - \xi ^{i^*} {\tilde
\epsilon}^+  \\
\delta \psi ^i = - \frac{i}{2} \partial _+ X ^ i \epsilon ^+
- {\tilde \epsilon }^- \Gamma ^i_{jk}\xi ^j\psi ^k &
\delta \psi ^{i^*}  =  \frac{i}{2} \partial _+ X^{i^*} \epsilon ^- +
 {\tilde \epsilon }^+ \Gamma ^{i^*}_{j^*k^*}\xi ^{j^*}\psi ^{k^*} \\
\delta \xi ^i  =  -\frac{i}{2} \partial _-X^i {\tilde \epsilon }^+
-\epsilon ^- \Gamma ^i_{jk}\psi ^j\xi ^k &
\delta \xi ^{i^*} = \frac{i}{2} \partial _-X^{i^*}{\tilde \epsilon }^-
+ \epsilon ^+ \Gamma ^{i^*}_{j^*k^*}\psi ^{j^*}\xi ^{k^*} \ .
\label{SSTRonshell}
\end{array}
\end{equation}
The supersymmetry algebra only closes on shell.

\subsection{Off shell formulation}
\noindent
To find
an off shell formulation, we introduce auxiliary fields $F^i,F^{i^*}$.
The action given above is then obtained by integrating out the
auxiliary fields in the following action~:
\begin{eqnarray}
S&=&-\frac{1}{2}g_{ij^*}(\partial _+X^i\partial _-X^{j^*}+\partial
_-X^i\partial _+X^{j^*})\nonumber\\
&&+ig_{ij^*}(\psi ^i\nabla _-\psi ^{j^*}+\psi ^{j^*}\nabla _-\psi
^i)\nonumber\\
&&+ig_{ij^*}(\xi ^i\nabla _+\xi ^{j^*}+\xi ^{j^*}\nabla _+\xi
^i)\nonumber\\
&&-F^iF^{j^*}g_{ij^*}+2\Gamma ^i_{jk}\xi ^j\psi
^kg_{ij^*}F^{j^*}+2F^ig_{ij^*}\Gamma ^{j^*}_{k^*l^*}\psi ^{l^*}\xi
^{k^*}\nonumber\\
&&-4\psi ^i\psi ^{j^*}\xi ^k\xi ^{l^*}
\partial _i\partial _{j^*}\partial _k\partial _{l^*}K\ ,
\end{eqnarray}
where $K$ is the K\"ahler potential with $g_{ij^*}=\partial _i\partial
_{j^*}K$. The $N=2$ susy transformation rules now become
\begin{equation}
\begin{array}{cc}
\delta X^i  =  \psi ^i \epsilon ^- + \xi ^i {\tilde \epsilon}^- &
\delta X^{i^*} = -\psi ^{i^*} \epsilon ^+ - \xi ^{i^*} {\tilde
\epsilon}^+  \\
\delta \psi ^i = - \frac{i}{2} \partial _+ X ^ i \epsilon ^+
- \frac{1}{2} F^i {\tilde \epsilon }^- &
\delta \psi ^{i^*}  =  \frac{i}{2} \partial _+ X^{i^*} \epsilon ^- -
  \frac{1}{2} F^{i^*}  {\tilde \epsilon }^+  \\
\delta \xi ^i  =  -\frac{i}{2} \partial _-X^i {\tilde \epsilon }^+
+\frac{1}{2} F^i \epsilon ^-  &
\delta \xi ^{i^*} = \frac{i}{2} \partial _-X^{i^*}{\tilde \epsilon }^-
+ \frac{1}{2} F^{i^*} \epsilon ^+  \\
\delta F^i = -i \partial_+ \xi^i  \epsilon ^+  + i \partial_- \psi^i
{\tilde \epsilon }^+ &  \delta F^{i^*}  = -i \partial_+ \xi^{i^*}
\epsilon ^-  + i \partial_- \psi^{i^*} {\tilde \epsilon }^- \ .
\label{SSTRoffshell}
\end{array}
\end{equation}
The twist we will perform is based on these transformation rules. They are
the same for any (rigid) $N=2$ theory in two dimensions. It is therefore
not surprising that the results of \cite{tlgqaq} can also be applied to
the case of sigma models.
The field equations of the auxiliary fields are
\begin{eqnarray} F^i&=&2\Gamma ^i_{jk}\xi ^j\psi ^k\nonumber\\
F^{i^*}&=&2\Gamma ^{i^*}_{j^*k^*}\psi ^{j^*}\xi ^{k^*}\ .
\end{eqnarray}
Using these field equations in the action and in the transformation rules,
one recovers the on shell formulation of the previous subsection.

\section{B-Twisting: the old approach}
\noindent
Following Witten \cite{ABTwist,tsm}, one can perform the B twist of the
$N=2$ model
by setting $\epsilon ^-={\tilde \epsilon }^-=0$. The other two supercharges
build up a spinless BRST operator $Q=G^++{\tilde G}^+$. All the fields
have zero spin, except $\psi ^i$ and $\xi ^i$, which have resp. spin -1 and
1.

\subsection{On shell twisting}
\noindent
For the on shell formulation, we get the BRST rules (acting from the left)~:
\begin{eqnarray}
\delta X^{i^*}=\psi ^{i^*}+\xi ^{i^*} &\qquad& \delta X^i=0\nonumber\\
\delta \xi ^{i^*}= \Gamma ^{i^*}_{j^*k^*}\psi ^{j^*}\xi ^{k^*} &\qquad&
\delta \xi ^i=-\frac{i}{2}\partial _-X^i\nonumber\\
\delta \psi ^{i^*}=- \Gamma ^{i^*}_{j^*k^*}\psi ^{j^*}\xi ^{k^*}&\qquad&
\delta \psi ^i=-\frac{i}{2}
\partial _+X^i \ . \label{BRSTonshell}
\end{eqnarray}
This BRST operator is nilpotent, and the action can be rewritten as
\begin{eqnarray}
S&=&4R_{ij^*kl^*}\psi ^i\psi ^{j^*}\xi ^k\xi ^{l^*}+ig_{ij^*}(\xi
^{j^*}-\psi ^{j^*})(\nabla _+\xi ^j-\nabla _-\psi ^j)\nonumber\\
&&+\delta [ -ig_{ij^*}( \psi ^i\partial _-
X^{i^*}+\xi ^i\partial _+X^{j^*})]\nonumber\\ &\equiv &S^0+\delta \Psi \ .
\label{TSMonshell}
\end{eqnarray}
First we assign ghost numbers to all the fields.
Imposing that the action has ghost number zero, and the BRST operator has
ghost number one, it follows
that $X^i$ and $X^{i^*}$ have ghost number zero, $\xi ^{i^*}$ and $\psi
^{i^*}$ have ghost number one and $\psi ^i$ and $\xi ^i$ have ghost number
minus one. With these assignements, the part $S^0$ of the action still
contains ghosts and antighosts, i.e. it is not the classical action,
which only depends
on the classical fields and not on the ghosts. Secondly, the fields
$X^i$ are merely lagrangian multipliers since they do not transform under
BRST. So the only classical field would be $X^{i^*}$, but then the
classical action should only depend on $X^{i^*}$.
It is clear that in this formulation the BRST interpretation is obscure.
We will now see that some of these
problems disappear when including the auxiliary sector.

\subsection{Off shell twisting}
\noindent
When the auxiliary fields are included, the BRST transformation rules are~:
\begin{eqnarray}
\delta X^{i^*}=\psi ^{i^*}+\xi ^{i^*} &\qquad& \delta F^i=i(\partial _+\xi
^i-\partial _-\psi ^i)\nonumber\\
\delta \xi ^{i^*}=\frac{1}{2}F^{i^*} &\qquad& \delta \xi
^i=-\frac{i}{2}\partial _-X^i\nonumber\\
\delta \psi ^{i^*}=-\frac{1}{2}F^{i^*} &\qquad& \delta \psi ^i=-\frac{i}{2}
\partial _+X^i \nonumber\\
\delta F^{i^*}=0 &\qquad& \delta X^i=0 \ . \label{BRSToffshell}
\end{eqnarray}
{}From these expressions it is obvious that $\delta^2=0$. It is proposed in
\cite{ABTwist} to interpret $\delta $ as a BRST operator of a so far
unspecified gauge symmetry. The action of the $\sigma $ model
can be written as
\begin{eqnarray}
S&=&
\delta [(F^i-2\Gamma ^i_{jk}\xi
^j\psi ^k)(\psi ^{j^*}-\xi ^{j^*})g_{ij^*} -ig_{ij^*}( \psi ^i\partial _-
X^{j^*}+\xi ^i\partial _+X^{j^*})]\nonumber\\ &\equiv &S^0+\delta \Psi \ .
\label{TSMofshell}
\end{eqnarray}
This is of the same form as a classical action $S^0$, supplemented by a
gauge
fixing action which is the BRST variation of a gauge fermion. The classical
action in this case is simply zero. There are two gauge fixing conditions.
One that restricts the $X$'s to be constant maps from the Riemann
surface $\Sigma $
to the target manifold ${\cal M}_K$. The other puts the auxiliary fields on
shell. The ghost numbers are the
same as in the previous subsection, and for the auxiliary fields one has
$gh(F^{i^*})=2=-gh(F^i)$. This means that they are ghosts for ghosts and
the gauge algebra is reducible. Indeed, there is a gauge symmetry
corresponding to arbitrary shifts in $X^{i^*}$. For this we have introduced
two ghosts instead of one, namely $\xi ^{i^*}$ and $\psi ^{i^*}$.
Whereas this redundancy somewhat complicates matters, it is
adequately handled by the ghosts for ghosts $F^{i^*}$.
When looking for an interpretation of the column on the right
in eqs.\ref{BRSToffshell}, there does arise a problem. It seems that it
can also be understood as
a reducible multiplet with $F^i$ as a classical field.
The form of its transformation rule leads to extra transformation on the
ghosts $\xi ^i$ and $\psi ^i$, and $X^i$ would be a multiplier. However,
this contradicts the ghost number assignments:
we can not interpret $F^i$ as a
classical field, since it has ghost number minus two. Analogous statements
hold for the would-be ghosts $\xi ^i$ and $\psi ^i$.

As we will see in the next section, all these
problems can be solved by defining a new BRST operator.

\section{A new formulation}
\noindent
To remedy this situation, {\it we propose to change the BRST operator}.
The previous BRST operator was obtained from the supersymmetries
with as BRST parameter $\Lambda=\epsilon^+ = \tilde \epsilon^+$.
Instead, we propose to use simply the first of these supersymmetries, and
interpret it as a BRST operator by itself.
The
second supersymmetry we propose to identify with the anti-BRST operator%
\footnote{For a review of
the use of BRST--anti-BRST symmetry of gauge theories, we refer to
\cite{BAB}.}%
. We will call these
operators $\bf s$ and $\bf{\bar s}$ respectively.
The transformation rules are:
\begin{equation}
  \begin{array}{ll}
   \bar {\bf s} X^{i*} =  \psi^{i*}  & {\bf s} X^{i*} = \xi^{i*} \\
   \bar {\bf s} \psi^i =  -\frac{i}{2} \partial _+ X^i & {\bf s} \psi^{i*}
   = -\frac{1}{2} F^{i*} \\
   \bar {\bf s} \xi^{i*} = \frac{1}{2} F^{i*}  & {\bf s} \xi^i =
-\frac{i}{2} \partial _- X^i \\
   \bar {\bf s} F^i = i \partial_+ \xi^i & {\bf s} F^i = -i \partial _-
   \psi^i \label{ouranti}\ ,
 \end{array}
\end{equation}
with all the other (anti)BRST transformations vanishing.
One easily
verifies the important nilpotency relations ${\bf s}^2 = {\bf {\bar s}}^2 =
{\bf s} {\bf {\bar
s}} + {\bf {\bar s} s} = 0$. Comparing with \eqn{BRSToffshell} we see that
the previous BRST operator  is the sum,
$\delta = \bf s + \bf{\bar s}$.
The invariance of the action under $s$ and $\bar s$ follows of course from
the original supersymmetries.
The condition that fixes the ghost number assignments
is now that $s$ raises the ghost number by one unit,  $\bar s$
lowers it by one unit, and the action has ghost number zero. All the bosons
have ghost number zero, and the fermions $\psi ^i,\psi ^{i^*},\xi ^i,\xi
^{i^*}$ have ghost numbers resp. 1,-1,-1,1.

With this new interpretation,
the  action of the $\sigma $ model can still be written
as the sum of a classical action and a gauge fixing part.
One easily computes
\begin{equation}
  S =  {\bf s}[-2ig_{ij^*}\partial _+X^{j^*}\xi ^i+2g_{ij^*}\psi
^{j^*}(F^i-2\Gamma ^i_{jk}\xi ^j\psi ^k)]\ .\label{TSMBAB}
\end{equation}
The classical part does not depend on $X^{i*}$,
and therefore one has a gauge (shift-)symmetry
$\delta X^{i*} = \varepsilon^{i*}$,
and the corresponding ghosts $\xi^{i*}$. In accordance with the spirit of
the BRST--anti-BRST scheme \cite{BAB}, one introduces also an
antighost $\psi^{i*}$, and its BRST variation $F^{i*}$.
Apart from this quartet, there is a second set of fields transforming
into each other, viz.  $F^i, \psi^i, \xi^i$ and $X^i$.
They ensure that one restricts the $X^i$ to be constant,
which was also the case in
the $\delta $ picture. Indeed, the BRST gauge
fixing condition ${\bf s}\xi ^i=-\frac{i}{2}\partial _-X^i$ forces these
maps to be holomorphic. It is then the anti-BRST operator that kills the
anti-holomorphic part of $X^i$, since it is anti-BRST exact.

There are still two things that are unsatisfactory. First, if $F^i$ is
interpreted as a classical field, and the classical action is zero, then
the gauge symmetry on $F^i$ would be an arbitrary shift. Looking at its
transformation rule we did not include this shift symmetry. Secondly
the identifications above do not yet exhibit the customary structure of
BRST-anti-BRST, which exhibits
more symmetry between ghosts and antighosts: the anti-BRST transformation
of the classical fields are usually
identical to their BRST transformation,
when replacing ghosts with antighosts. This is not the case for the
second set of fields above,
since we then also have to interchange $\partial _+$ and $\partial _-$.

In the $N=2$ $\sigma $-model,  the starred and unstarred fields occur
symmetrically.
The twist has lifted this symmetry: the former are all spinless, but
$\psi ^i$  and $\xi ^i$ have helicities 1 and -1 respectively.
One can construct $\psi ^idx^+$ and $\xi ^idx^-$, which behave as one
forms under holomorphic coordinate transformations. The asymmetry is
mirrored in
the derivatives in the transformation laws for the second set, which is in
accordance with the helicity-assignment. At the same time, one can also
consider
$F^i$ to be a two-form, which can not be distinguished from a scalar
in the treatment with a flat metric.
The BRST--anti-BRST symmetry can be redressed by the
following non-local change of field variables:
\begin{eqnarray}
\psi ^i&=&\partial _+\chi ^i\nonumber\\
\xi ^i&=&\partial _-\rho ^i\nonumber\\
F^i&=&\partial _-\partial _+H^i  \ .
\end{eqnarray}
All the fields on the right hand side are scalars.
Remark that the Jacobian of this transformation is
equal to unity, at least formally, since the
contributions from the fermions cancel against the bosons.
For the new variables we can take the transformation rules
\begin{eqnarray}
\bar {\bf s} \chi^i= -\frac{i}{2}X^i &
\qquad& {\bf s} \rho^i =
-\frac{i}{2}X^i \nonumber\\
\bar {\bf s} H^i = i \rho^i &\qquad& {\bf s} H^i = -i\chi^i \ ,
\label{qaq for nonlocal}
\end{eqnarray}
to reproduce the so far ``unexplained'' rules in (\ref{ouranti}).
They now correspond to a shift symmetry for the field $H^i$, introducing
the ghost field $\chi^i$. The antighost is $\rho^i$, and $X^i$
completes the quartet.
It is clear that we  have uncovered a manifest BRST anti-BRST symmetry.
The action, when written in terms of the new fields, is of course still
BRST exact: one simply writes the exact term in \eqn{TSMBAB} in terms of
the new variables.

This allows the following interpretation. One starts from two classical
fields, $X^{i^*}$ and $H^{i}$.
The classical action is zero, and the symmetries are shift symmetries,
with ghosts $\xi ^{i*}$ and $\chi ^i$.
Then one introduces  antighosts  $\psi ^{i^*}$ and $\rho ^i$, and
Lagrange multipliers  $X^i$ and $F^{i^*}$. This completes the field
content of the theory.
Note that the actual content of the resulting TFT depends heavily on
the gauge fixing procedure, as usual: there are no physical local
fluctuations, but global variables may remain.

Having changed the BRST operator, we now discuss the implications of this
change. First of all, we investigate whether we still have a topological
theory in the sense that the energy-momentum is BRST exact for the new BRST
operator. Afterwards, we will investigate whether the physical content
(observables) of the theory has changed.

\section{The energy momentum tensor}
\noindent
There are
two metrics in the model~: the world sheet metric $h_{\alpha \beta }$,
which
was taken to be flat, and the space time K\"ahler metric $g_{ij^*}$. The
world sheet metric is external. The space time metric is a function of
$X^i$ and $X^{i^*}$, which are integration variables in the path integral.
We can
thus only study the dependence of the path integral on the $h_{\alpha
\beta }$ metric, by computing the energy momentum tensor.
The computation is analogous to the
Landau-Ginzburg model \cite{tlgqaq} and we find
\begin{eqnarray}
     T^B_{++}&=&-g_{ij^*}\partial_+ X ^{j^*} \partial_+ X^i + 2i
           g_{ij^*}\psi^i
     \nabla _+\psi^{j*}=\bar {\bf s} \left[ 2i g_{ij^*}\psi^i \partial_+ X
             ^{j^*} \right] \nonumber \\
     T^B_{--}&=&-g_{ij^*} \partial_- X ^{j^*} \partial_- X^i + 2i
                g_{ij^*}\xi^i
     \nabla _- \xi^{j*}={\bf s} \left[ 2ig_{ij^*} \xi^i \partial_- X
            ^{j^*} \right] \nonumber\\
     T^B_{+-}&=&0 \ .
\end{eqnarray}
After the derivation, we have taken the metric to be flat. These are
therefore the relevant operators for variations of correlation functions
around a flat metric. We see that, although the action is BRST exact, the
(++) component of the energy momentum tensor is only anti-BRST exact. This
is because the BRST operator depends on the metric and one cannot commute
the BRST variation and the derivative w.r.t. the metric \cite{tftbv}.

To prove metric independence of correlation functions, one needs
not only BRST invariance, but also
the Ward identity for the anti-BRST operator.
What is needed is that the physical operators are
BRST invariant, {\em  and} that their anti-BRST variation is BRST
exact. For a more complete argument, see \cite{tlgqaq}.

The proper cohomological formulation is, that one first determines the
${\bf s}$ cohomology, a space of equivalence classes.
The operator ${\bar {\bf s}}$ is well defined and
nilpotent in that space, so that {\em its} cohomology can be used
as our characterisation of physical states%
\footnote{This procedure is  more familiar than it sounds. When one
considers a BRST cohomology class, it is quite common that it contains
only one anti-BRST invariant member (up to a factor) --- in which case
that member is considered to be the physical one. See for example
\cite{Hwang}.}.
This characterisation is not arbitrary, but more or less forced upon us
by the requirement that the energy momentum tensor is trivial.%
We now investigate this cohomology.

\section{The spectrum}
\noindent
The observables of the B twisted $\sigma $ model were first computed in
\cite{ABTwist}. This was done in the on shell formulation, i.e. without
using the auxiliary fields. In order to compare with the off shell
formulation, one has to compute the weak cohomology of the BRST operator.
In this way one eleminates the auxiliary fields again.
For topological LG models, this leads to dividing out the vanishing
relations
$\kappa \partial _iW=0$, where $W$ is the LG potential, since it is weakly
(using the field equations of the auxiiary fields) equal to the BRST
variation of $\psi ^{i^*}$. The
local (zero forms) observables are then the elements of the chiral ring.
We will see below to what it will lead in the case of the $\sigma $ model.

\subsection{The $\delta $ cohomology}
\noindent
Let us start with the $\delta $ cohomology. One can make the field
redefinitions $c^{i^*}=\psi ^{i^*}+\xi ^{i^*}$ and ${\bar c}^{i^*}=\xi
^{i^*}-\psi ^{i^*}$. The BRST algebra on the new fields is then
\begin{eqnarray}
\delta X^{i^*}=c^{i^*} &\qquad& \delta {\bar c}^{i^*}=F^{i^*}\nonumber\\
\delta c^{i^*}=0 &\qquad & \delta F^{i^*}=0\ ,\label{starredBRST}
\end{eqnarray}
together with the unstarred sector in the left column of
\eqn{BRSToffshell}.

At first sight, one might think that there is no cohomology in the starred
sector, since $\{X^{i^*},c^{i^*}\}$ and $\{{\bar c}^{i^*},F^{i^*}\}$ form
trivial pairs that drop out of the cohomology. For two reasons this is not
true.
A first reason is that we are interested in the weak cohomology.
The field equations imply relations between the different fields, so that
the usual reasoning for eliminating trivial pairs does not apply.
Going to the basis with \cite{ABTwist}
\begin{equation}
{\bar c}_i=g_{ij^*}{\bar c}^{j^*}\ ,
\end{equation}
one finds for the second pair $\{{\bar c}^{i^*},F^{i^*}\}$
\begin{equation}
\delta {\bar c}_i=-y_{F^i}=g_{i k^*}( F^{k^*} -\partial
_{j^*}g^{k^*l}c^{j^*}{\bar c}_l)\ ,
\end{equation}
where $y_{F^i}$ stands for the field equation of $F^i$. In this way, the
auxiliary fields are eleminated from the spectrum. Therefore,
${\bar c}_i$ is weakly
BRST invariant. Since it is not exact,  it remains in the weak cohomology.

Consider now a section $V^{(0,p)}_{(q,0)}\in \Omega ^{(0,p)}\times \wedge
^qT^{(1,0)}{\cal M}_K$, i.e. $V^{(0,p)}_{(q,0)}$ takes the form
\begin{equation}
V^{(0,p)}_{(q,0)}=dX^{i^*_1}\wedge ... \wedge
dX^{i^*_p}V^{j_1...j_q}_{i^*_1...i^*_p}\frac{\partial }{\partial
X^{j_1}}\wedge ... \wedge \frac{\partial }{\partial X^{j_q}}\ .
\end{equation}
We can associate a zero form to any such section
\begin{equation}
{\cal O}[V^{(0,p)}_{(q,0)}]
=c^{i^*_1}...c^{i^*_p}V^{j_1...j_q}_{i^*_1...i^*_p}{\bar c}_{j_1}...
{\bar c}_{j_q}\ .
\end{equation}
{}From the identity
\begin{equation}
\delta {\cal O}[V^{(0,p)}_{(q,0)}]\approx {\cal O}[{\bar \partial
}V^{(0,p)}_{(q,0)}]\
\end{equation}
it follows that the BRST cohomology and the twisted
Dolbeault cohomology on the target manifold are isomorphic
(see \cite{ABTwist,Freboek} for more details):
 we have that ${\cal O}[V^{(0,p)}_{(q,0)}]$ is BRST invariant
if ${\bar \partial }V^{(0,p)}_{(q,0)}=0$ and BRST exact if $V^{(0,p)}_{(q,0)}=
{\bar \partial }S^{(0,p-1)}_{(q,0)}$. This isomorphism between BRST
cohomology and twisted Dolbeault cohomology only holds for the local
observables, i.e the zero forms. To know the global observables, one must
solve the descent equations, see \cite{ABTwist}.

{}From this isomorphism one can see the second reason why the $\delta$
cohomology may not be trivial, namely because of global
properties of the target manifold. While one can locally write any closed
form as an exact differential, one cannot do this in a global way.
For the BRST cohomology this means that functions of
$\{X^{i^*}, c^{i^*}\}$ may not drop out of the spectrum.

\subsection{The ${\bar {\bf s}}$ in ${\bf s}$ cohomology}
\noindent
In \cite{tlgqaq} we, have shown the equivalence of the $\delta $ cohomology
with the ${\bar {\bf s}}$ in ${\bf s}$ cohomology in
topological Landau-Ginzburg models. Here, we want to argue that this
equivalence also holds for B--twisted topological $\sigma $ models. We will
only indicate where the proof differs from the one given in \cite{tlgqaq}.
In the unstarred sector, everything goes through as in \cite{tlgqaq}.
In the starred sector,  we have a non zero
$\delta $ cohomology.
The $\{c^{i^*},{\bar c}_i\}$ basis of the previous subsection is not
convenient anymore. This is because of the chiral split we have made when
defining the BRST anti-BRST complex.
Instead we will define
\begin{equation}
{\bar \psi }_i=g_{ij^*}\psi ^{j^*}\ .
\end{equation}
This leads to
\begin{eqnarray}
{\bf s}X^{i^*}=\xi ^{i^*} &\qquad& {\bar {\bf s}}X^{i^*}=g^{i^*j}{\bar \psi
}_j\nonumber\\
{\bf s}\xi^{i^*}=0 &\qquad& {\bar {\bf s}}\xi ^{i^*}\approx
\partial _{k^*}g^{i^*j}{\bar \psi }_j\xi ^{k^*} \nonumber\\
{\bf s}{\bar \psi }_i=\frac{1}{2}y_{F^i}
=-1/2 g_{i l^*} (F^{l^*}+ \partial _{k^*}g^{l^*j}{\bar \psi }_j\xi ^{k^*})\
\approx 0 &\qquad& {\bar {\bf
s}}{\bar \psi }_i=0 ,
\end{eqnarray}
together with the unstarred sector, see \eqn{ouranti}.

To compute the ${\bf s}$ cohomology, we take again a section
$V^{(0,p)}_{(q,0)}
\in \Omega ^{(0,p)}\times \wedge ^qT^{(1,0)}{\cal M}_K$.
Now, we associate with it the zero form operator
\begin{equation}
{\cal O}[V^{(0,p)}_{(q,0)}]
=\xi^{i^*_1}...\xi^{i^*_p}V^{j_1...j_q}_{i^*_1...i^*_p}{\bar \psi
}_{j_1}... {\bar \psi }_{j_q}\ . \label{OV}
\end{equation}
One finds again that
\begin{equation}
{\bf s}{\cal O}[V^{(0,p)}_{(q,0)}] \approx{\cal O} [{\bar
\partial }V^{(0,p)}_{(q,0)}] \ .
\end{equation}
This means that the ${\bf s}$ cohomology is also isomorphic to the twisted
Dolbeault cohomology, and thus to the $\delta $ cohomology,
at least for the zero forms (for higher forms, the reasoning may be
extended using descent equations).
This indicates that the second step in the computation  of
the ${\bar {\bf s}}$ in the ${\bf s}$ cohomology does not give further
restrictions. To check
this, we take the anti-BRST variation of an element of the ${\bf s}$
cohomology and find
\begin{equation}
{\bar {\bf s}}{\cal O}[V^{(0,p)}_{(q,0)}] \approx
{\bf s}{\cal O}[V^{(0,p-1)}_{(q+1,0)}] \ ,
\end{equation}
where ${\cal O}[V^{(0,p-1)}_{(q+1,0)}]$
can be read off from \eqn{OV} raising an index using the
metric, more explicitly:
\begin{equation}{\cal O}=
(-)^{q-1}\,p\, [\xi^{i^*_1}...\xi^{i^*_{p-1}}
V^{j_1...j_q}_{i^*_1...i^*_{p-1}i^*_p}
g^{i^*_pj_{q+1}}{\bar \psi}_{j_1}...{\bar \psi}_{j_{q+1}}] .
\end{equation}
This means that, for zero forms, the ${\bar {\bf s}}$ operator
in the ${\bf s}$ cohomology is equal to zero, and
the ${\bar {\bf s}}$ cohomology in the ${\bf s}$
cohomology  is equivalent to the $\delta $ cohomology.

\vspace{3mm}
\noindent{\bf Acknowledgement}:
The work of FDJ was supported by the Human Capital and Mobility Programme
by the network on {\it Constrained Dynamical Systems}. WT acknowledges the
financial support of his employer, the N.F.W.O., Belgium.

\end{document}